\documentstyle[12pt,fleqn]{article}
\addtocounter{section}{-1}
\begin{document}

\begin{center}

\vspace{2cm}

{\large {\bf {
Electric capacitance as nanocondensers\\
in zigzag nanographite ribbons and zigzag carbon nanotubes
} } }

\vspace{1cm}

{\rm Kikuo Harigaya\footnote[1]{E-mail address: 
\verb+k.harigaya@aist.go.jp+; URL: 
\verb+http://staff.aist.go.jp/k.harigaya/+}}

\vspace{1cm}

{\sl Nanotechnology Research Institute, AIST, 
Tsukuba 305-8568, Japan}\footnote[2]{Corresponding address}\\
{\sl Synthetic Nano-Function Materials Project, 
AIST, Tsukuba 305-8568, Japan}

\vspace{1cm}

(Received~~~~~~~~~~~~~~~~~~~~~~~~~~~~~~~~~~~)
\end{center}

\vspace{1cm}

\noindent
{\bf Abstract}\\
Electronic states in nanographite ribbons with zigzag edges
and zigzag carbon nanotubes
are studied using the extended Hubbard model with nearest 
neighbor Coulomb interactions.  The nearest Coulomb interactions 
stabilize electronic states with the opposite electric 
charges separated and localized along both edges.  Such 
states are analogous as nanocondensers.  Therefore, 
electric capacitance, defined using a relation of
polarizability, is calculated to examine nano 
functionalities.  We find that the behavior of the 
capacitance is widely different depending on whether 
the system is in the magnetic or charge polarized phases.  
In the magnetic phase, the capacitance is dominated by 
the presence of the edge states while the ribbon width is small.  
As the ribbon becomes wider, the capacitance remains with 
large magnitudes as the system develops into metallic 
zigzag nanotubes.  It is proportional to the inverse 
of the width, when the system corresponds to the semiconducting 
nanotubes and the system is in the charge polarized phase also.
The latter behavior could be understood by the presence of 
an energy gap for charge excitations.

\noindent
keywords: electronic states, capacitance, graphite ribbons, carbon nanotubes

\pagebreak

\section{Introduction}

Nano carbon materials have been attracting much attention 
both in the fundamental science and in the interests of
application to nanotechnology devices.$^{1,2)}$  Their 
physical and chemical natures change variously depending 
on geometries.$^{1-3)}$  In carbon nanotubes, diameters 
and chiral arrangements of hexagonal pattern on tubules 
decide whether they are metallic or not.$^{1,2)}$  In 
addition, carbon systems with open boundaries show 
unusual features.

In nanographites, the peripheral edges strongly affect 
the electronic states,$^{3)}$ and the nanographite ribbon 
with zigzag edges has the nonbonding molecular orbitals 
localized mainly along the zigzag edges. The partly 
flat electronic bands give rise to a sharp peak in the 
density of states.  The on-site Coulomb interaction by the 
Hubbard model has suggested to induce the spin polarized (SP)
state, and the most spin density concentrates along the 
zigzag edges forming the ferrimagnetic orders 
due to the edge states.  As to the effects of the Coulomb 
interactions, another qualitatively important problem has
come out.  This is by no means the competition between 
the spin and charge orderings in the nanographite system 
due to the on-site and nearest neighbor Coulomb interactions. 
In fact, for two dimensional (2D) boundless graphite, 
the possibility of the stable charge density wave (CDW) 
state compared with the spin density wave (SDW) state 
was reported.$^{4,5)}$

Recently,$^{6)}$ we have demonstrated that the nearest neighbor 
Coulomb interaction stabilizes a novel charge polarized 
(CP) state with a finite electric dipole moment in 
zigzag ribbons, and that it competes with the SP state.
Such the CP state is caused by the interplay 
between the nearest neighbor Coulomb interaction and 
the edge states in nanographene ribbons with zigzag 
edges using the extended Hubbard model.  This study
has been treated rather formally by changing the 
interaction parameters in the wide ranges.  Though it
has been discussed that the transverse electric field 
might induce the first order phase transition from the 
SP state to the CP state,$^{6)}$ we need further study 
in order to reveal what roles such the novel
SP and CP states play in actual physical quantities
measurable in experiments.  Such theoretical studies
should be useful when materials specific properties
due to edge localized states can be expected.

In this paper, electric capacitance is calculated extensively
in order to examine possible functionalities as 
``nano-size condensers" of nanographite materials.
The positive and negative charge accumulations along
the edges of the both sides have a pattern analogous
to a condenser, if each set of edge site atoms is regarded
as one electrode.  One of them is the positive electrode, 
and the other is the negative one.  The spacer between
two electrodes is the inner part of the zigzag nanoribbon.
The capacitance is calculated with the relation of the
charge polarizability.  This definition is the usual
capacitance for the SP state because there is not
charge polarization without an electric field.  However, 
the definition becomes different for the CP state, 
because the self-polarized charge is present at the edge 
sites.  In this case, the calculated capacitance is 
a ``differential capacitance" as defined by the relation 
of polarizability.

The calculated electric capacitance will be discussed
with scopes of nano functionalities.  In the magnetic phase, 
the capacitance is dominated by the presence of the edge 
states while the ribbon width is small.  As the ribbon 
becomes wider, the capacitance remains with large magnitudes 
as the system develops into metallic zigzag nanotubes, 
while it is proportional to the inverse of the width 
when the system corresponds to the semiconducting 
nanotubes.  In the charge polarized phase, the capacitance 
is inversely proportional to the graphite width.

This paper is organized as follows.  In section 2, we explain
the extended Hubbard model.  In section 3, we give a brief
review of the SP and CP states in order to look at 
electronic properties.  In section 4, calculated electric 
capacitance is reported and discussion is given.
The paper is closed with a short summary in section 5.

\section{Model and method}

Figure 1 illustrates the geometry of a zigzag ribbon 
with an inversion symmetry.  Here, $N$ and $L$ are the 
width and length of the ribbon, respectively.  The dashed 
rectangle denotes a unit cell.  The periodic boundary condition 
is set along the $y$ axis of Fig. 1.  Since the 
zigzag ribbon is bipartite, $A$ and $B$ sites are assigned 
by filled and open circles, respectively.  All the 
twofold coordinated sites in the lower and upper 
zigzag edges belong to the $A$ and $B$ sublattices, 
respectively, at which edge states are mainly localized.

We consider the $\pi$-electron system on the 
zigzag ribbon using the extended Hubbard Hamiltonian 
with the on-site $U$ and nearest-neighbor $V$ Coulomb 
interactions.  The model is as follows:
\begin{eqnarray}
H &=& - t \sum_{\langle i,j \rangle,\sigma}
( c_{i,\sigma}^\dagger c_{j,\sigma} + {\rm h.c.} )
+ U \sum_{i} 
(c_{i,\uparrow}^\dagger c_{i,\uparrow} - \frac{n_{\rm el}}{2})
(c_{i,\downarrow}^\dagger c_{i,\downarrow} 
- \frac{n_{\rm el}}{2}) \nonumber \\
&+& V \sum_{\langle i,j \rangle}
(\sum_\sigma c_{i,\sigma}^\dagger c_{i,\sigma} - n_{\rm el})
(\sum_\tau c_{j,\tau}^\dagger c_{j,\tau} - n_{\rm el}),
\end{eqnarray}
where $t$ ($> 0$) is the hopping integral between 
the nearest neighbor $i$th and $j$th sites; the sum with
$\langle i,j \rangle$ is taken for all the pairs 
of the nearest neighbor sites; $c_{i,\sigma}$ annihilates 
an electron of spin $\sigma$ at the $i$th site; $n_{\rm el}$ 
is the average electron density of the system.

We adopt the standard Unrestricted Hartree-Fock approximation 
to this model:$^{3,6,7)}$
\begin{equation}
c_{i,\uparrow}^{\dagger}c_{i,\uparrow} 
 c_{i,\downarrow}^{\dagger}c_{i,\downarrow} 
\Rightarrow \langle c_{i,\uparrow}^{\dagger}c_{i,\uparrow}\rangle
c_{i,\downarrow}^{\dagger}c_{i,\downarrow}
+c_{i,\uparrow}^{\dagger}c_{i,\uparrow}
\langle c_{i,\downarrow}^{\dagger}c_{i,\downarrow}\rangle
-\langle c_{i,\uparrow}^{\dagger}c_{i,\uparrow}\rangle
\langle c_{i,\downarrow}^{\dagger}c_{i,\downarrow}\rangle.
\end{equation}

\section{Magnetic and charge ordered states}

In this section, we briefly review the competition between
the SP and CP states, in order to give an introduction to
the materials and their electronic properties.

Figure 2 (a) shows the charge density distribution in 
the CP state with $N \times L=4 \times 20$, where $U=0.3t$ 
and $V=0.4t$, $t$ being the hopping integral between 
neighboring carbons.  For comparison, Fig. 2 (b) shows 
the spin density profile of the SP state for $U=1.0t$ and $V=0$. 
The CP state has no spin density at every site, while the 
SP state has no charge density at every site.  In the CP 
state, the upper zigzag edge is charged positively, while 
the lower zigzag edge is charged negatively.  The 
distribution pattern of the charge density in the CP state 
is just like that of the spin density in the SP state. The 
charge polarization in the CP state can be explained by the 
interplay between the edge states and the Fermi-instability 
of the flat bands due to the interaction $V$.  
The CP state has a finite electric dipole moment pointing 
from the upper edge to the lower edge along the $x$-axis of Fig. 1. 
It is formed in the homogeneous system and breaks 
the inversion symmetry.

The relative stabilities between the CP and SP states have
been investigated and summarized in the phase diagram.$^{6)}$
While $U$ is fairly larger than $V$, the SP state is
more stable.  As $V$ becomes stronger, the system exhibits
the first order phase transition from the SP state
to the CP state.  The charge order of Fig. 2 (a) occurs 
in the CP phase, while the spin order of Fig. 2 (b) 
in the SP phase.

\section{Electric capacitance}

In order to examine nano functionalities
as``nano-size condensers", electric capacitance of the 
nanographite ribbons is calculated. The idea of the
calculation is shown schematically in Fig. 3.  We assume 
that the two sets of carbon atoms at the zigzag edges are 
regarded as positive and negative electrodes, respectively.  
The absolute value of the net variation of the accumulated 
charge is divided by the strength of the small applied 
voltage, and the capacitance is obtained.  In Fig. 3 (b),
the quantity $Q_0$ is the net charge over the $L/2$ carbon 
atoms at the zigzag edge sites, when the static electric field
is absent.  The part $dQ$ is the change of the net 
charge with respect to the small field which is parallel
to the $z$-axis as shown in Fig. 3 (a).  The capacitance 
$C$ is calculated using the relation of polarizability
$dQ = C dV$, where $dV$ is the change of voltage due to 
the static electric field. The similar method has been used
for the calculation of the capacitance using the first
principle method.$^{8)}$  We take $dV=0.01t$ and $t=2.5$ eV.
We also assume that the bond length between carbons is 1.45~\AA.

Figure 4 shows the calculated results for the system in
the SP state.  The actual magnitudes in the
logarithmic scale are plotted in Fig. 4
(a) against the ribbon width in the scale of~\AA, and 
their inverse values are shown in (b).  When the width
$N$ of the system increases, it develops into the $(L/2,0)$ 
zigzag nanotube due to the periodic boundary condition along 
the $y$ direction of Fig. 1.  The nanotube is metallic when $L/2$ is 
a multiple of three, and is semiconducting for others$^{2)}$
in the noninteracting model.  While the 
ribbon width is small (approximately $< 10$~\AA), the 
capacitance would be dominated by the presence of the edge 
states.  Even though the ribbon width becomes larger,
the capacitance remains with large magnitudes for $L=18$,
as shown by squares in Fig. 4 (a).  This would be related
with the resulting metallic properties of the long enough
$(9,0)$ nanotube.  On the other hand, the capacitance for 
$L=20$ is proportional to the inverse width for larger widths,
as shown in Fig. 4 (b).  The presence of the semiconducting
gap in the $(10,0)$ nanotube might result in the distinctive 
difference.

Figure 5 summaries the calculated capacitance for the 
system in the CP state.  The two sets of $L=18$ and 20
are shown, too.  We find that the capacitance is inversely 
proportional to the graphite width [Fig. 5 (b)].  This 
behavior does not depend on whether the long enough system 
is the metallic (9,0) or semiconducting (10,0) nanotube.  
As the system has the charge orders, electrons would
become less mobile than those in the system without ordered
states.  The behavior could be understood by the presence 
of an energy gap for charge excitations.

We look at interaction parameter dependences of the calculated
capacitance.  Figures 6 and 7 show the results for the ribbon 
length $L=18$ and 20, respectively.  When the ribbon width becomes
wider, the system develops into (9,0) and (10,0) zigzag nanotubes.
The former is metallic and the latter is semiconducting.
The Coulomb interaction strengths are $(U,V)=(1.0t,0)$ and 
$(2.0t,0)$ for the SP state, while $(U,V)=(0,0.5t)$ and 
$(0,0.6t)$ for the CP state.  As the interaction strength
$U$ increases, the spin orderings become stronger.
This gives rise to a stiff response to the applied field,
thus the magnitude of $C$ decreases in the SP state.
Similarly, the magnitude of $C$ decreases in the CP state,
as the charge orderings become stronger with the increase
of the parameter $V$.

The intrinsic capacitance of the single wall carbon 
nanotube can be estimated as follows.  We consider a
nanotube with finite length which is present in a 
quantum box of the length $L_y$.  The wavenumber
is quantized with the interval $\Delta k = 2\pi/L_y$.
Addition energy of an electron to the nanotube in the
quantum box can be equated with an energy as a dielectric
system.  The addition energy can be estimated to be $\Delta E 
=(h/2\pi) v_F \Delta k \cdot (1/2) \cdot (1/2)$, where 
$v_F$ is the Fermi velocity, the first factor (1/2) comes from 
the degeneracies of spin, and the second one (1/2) is due
to the degeneracy of electronic states near the Fermi
energy.  Here, we assume that the nanotube interacts
with circumstances, and the degeneracies of spin
and electronic states are lifted slightly. Therefore, $\Delta E$
is the overall average level spacing including spin
and number of electronic bands.  By setting 
$\Delta E = e^2/ 2 C_Q L_y$, we obtain the quantum 
capacitance per unit length: $C_Q = 2 e^2/h v_F$.
For carbon nanotubes $v_F = 8 \times 10^5$ (m/s),$^{9)}$ 
so that $C_Q = 100$ (aF/$\mu$ m) $= 10^{-20}$ (F/\AA).
In fact, the capacitance per unit length has been
measured to be 190 (aF/$\mu$ m)$^{10)}$ for example,
and the above rough estimation explains the experimental
magnitude fairly well.  Such the order of magnitudes 
of the estimation and experimental value also agrees 
with that of the capacitance obtained in the calculation
within the difference of a few order of magnitudes.
Even though the detailed definitions of the capacitances
are different mutually, the resulting order of magnitudes
should reflect quantum characters of electronic systems.

We can give a reason of the obtained magnitudes of the
capacitance by analogy with a classical nano-size
condenser, too.  The calculated order of magnitudes of 
the capacitance around $10^{-20} \sim 10^{-19}$ F might 
be estimated as follows.  When we consider a classical 
parallel-electrode condenser with the area of the 
electrode $S$, the capacitance is given by $C = \epsilon (S/d)$, 
where $\epsilon$ is a dielectric constant of the spacer, 
and $d$ is the distance between the electrodes.  We shall 
assume that $S$ for the nanographite ribbon is estimated 
as (the effective thickness of the graphite plane 
$\sim$ the interlayer distance of the stacked graphite 
$\sim$ 3.5~\AA) $\times$ (the length of the ribbon).
We obtain the magnitude $C \sim 3 \times 10^{-20}$
F for $N=2$, $L=20$, and $\epsilon$ of the empty space.  
Though uncertainties of a few factors in $\epsilon$ will be
present as a material constant, the similar order 
of magnitudes would be obtained.  Therefore, the 
calculated order of magnitudes seems reasonable for 
nano-size systems.

The inversely proportional behavior against the width
might indicate that electric field between the two
electrodes is confined within the graphite plane, and 
does not come out to the outside. If the electric field
comes out and the each electrode can be regarded as a
metallic wire, the capacitance would show a logarithmic
dependence as a function of the ribbon width.

\section{Summary}

In nanographite ribbons with zigzag edges,$^{6)}$ the charge-polarized
state is present.  This state has a finite electric dipole 
moment caused by the interplay between the nearest neighbor 
Coulomb interaction and the edge states, and competes with 
the spin-polarized state caused by the on-site Coulomb 
interactions.

In this paper, electric capacitance has been calculated in order
to test the nano functionalities.  In the magnetic phase, 
the capacitance is dominated by the presence of the edge 
states while the ribbon width is small.  As the ribbon 
becomes wider, the capacitance remains with large magnitudes 
as the system develops into metallic zigzag nanotubes, 
while it is proportional to the inverse of the width 
when the system corresponds to the semiconducting 
nanotubes.  In the charge polarized phase, the capacitance 
is inversely proportional to the graphite width.

\begin{flushleft}
{\bf Acknowledgments}
\end{flushleft}

This work has been supported partly by Special Coordination 
Funds for Promoting Science and Technology, and by NEDO 
under the Nanotechnology Program.

\pagebreak
\begin{flushleft}
{\bf References}
\end{flushleft}

\noindent
$1)$ S. Iijima, Nature {\bf 354}, 56 (1991).\\
$2)$ R. Saito, G. Dresselhaus, and M. S. Dresselhaus, 
``Physical Properties of Carbon Nanotubes'', 
(Imperial College Press, London, 1998).\\
$3)$ M. Fujita, K. Wakabayashi, K. Nakada, 
and K. Kusakabe, J. Phys. Soc. Jpn. {\bf 65}, 1920 (1996).\\
$4)$ F. R. Wagner and M. B. Lepetit, 
J. Phys. Chem. {\bf 100}, 11050 (1996).\\
$5)$ A. L. Tchougreeff and R. Hoffmann, 
J. Phys. Chem. {\bf 96}, 8993 (1992).\\
$6)$ A. Yamashiro, Y. Shimoi, K. Harigaya, and K. Wakabayashi,
Phys. Rev. B {\bf 68}, 193410 (2003).\\
$7)$ K. Wakabayashi and K. Harigaya, 
J. Phys. Soc. Jpn. {\bf 72}, 998 (2003).\\
$8)$ N. Nakaoka and K. Watanabe, Eur. Phys. J. D {\bf 24}, 397 (2003).\\
$9)$ S. G. Lemay {\sl et al.}, Nature {\bf 412}, 617 (2001).\\
$10)$ T. D\"{u}rkop, S. A. Getty, E. Cobas, and
M. S. Fuhrer, Nano Lett. {\bf 4}, 35 (2004).

\pagebreak

\begin{flushleft}
{\bf Figure Captions}
\end{flushleft}

\mbox{}

\noindent
Fig. 1. Schematic structure of the nanographite ribbon 
with zigzag edges (zigzag ribbon).  The filled and open 
circles are $A$ and $B$ sites, respectively.
The dashed rectangle denotes a unit cell.

\mbox{}

\noindent
Fig. 2.  (a) Charge density distribution of the 
charge-polarized (CP) state for $U=0.3t$ and $V=0.4t$, 
and (b) the $z$ component of spin density distribution 
of the spin-polarized (SP) state for $U=t$ and $V=0$, 
on a zigzag ribbon with $4 \times 20$ sites.  The open 
and filled circles show positive and negative densities.  
Their radii are proportional to the magnitudes of the 
charge or spin densities: the maximum is 0.18 in (a) 
and 0.12 in (b).

\mbox{}

\noindent
Fig. 3. (a) Periodic nanoribbon.  Periodic boundary
condition is applied to the one-dimensional direction of
the nanoribbon.  The nanoribbon strip is perpendicular
to the $x$-$y$ plane.  Small static electric
field is parallel to the $z$-axis. (b) Analogy with
a parallel electrode condenser.  The quantity $Q_0$
is the net charge over the $L/2$ carbon atoms at the
zigzag edge sites, when the static electric field
is absent.  The part $dQ$ is the change of the net 
charge with respect to the small field.  The capacitance 
$C$ is obtained using the relation of polarizability
$dQ = C dV$, where $dV$ is the change of voltage
due to the static electric field.

\mbox{}

\noindent
Fig. 4.  The electric capacitance calculated 
for the SP state at $(U,V)=(1.0t,0)$. Two sets of the 
ribbon lengths $L=18$ (squares) and 20 (circles) are 
considered.  The magnitude of the capacitance (a) and 
its inverse (b) are plotted against the ribbon width 
in the scale of~\AA.

\mbox{}

\noindent
Fig. 5. The electric capacitance calculated 
for the CP state at $(U,V)=(0,0.5t)$.  Two sets of the 
ribbon lengths $L=18$ (squares) and 20 (circles) are considered.  
The magnitude of the capacitance (a) and its inverse (b) 
are plotted against the ribbon width in the scale of~\AA.

\mbox{}

\noindent
Fig. 6. The electric capacitance calculated for the ribbon 
length $L=18$.  The Coulomb interaction strengths are
$(U,V)=(1.0t,0)$ and $(2.0t,0)$ for the SP state,
while $(U,V)=(0,0.5t)$ and $(0,0.6t)$ for the CP state. 
The magnitude of the capacitance (a) and its inverse (b) 
are plotted against the ribbon width in the scale of~\AA.

\mbox{}

\noindent
Fig. 7. The electric capacitance calculated for the ribbon 
length $L=20$.  The Coulomb interaction strengths are
$(U,V)=(1.0t,0)$ and $(2.0t,0)$ for the SP state,
while $(U,V)=(0,0.5t)$ and $(0,0.6t)$ for the CP state. 
The magnitude of the capacitance (a) and its inverse (b) 
are plotted against the ribbon width in the scale of~\AA.

\end{document}